\documentstyle[pra,aps,preprint]{revtex}
\tightenlines
 \begin{document}
 \draft
 \title{Cosmic Coincidence with a new Type of Dark Matter}

\author{E. I. Guendelman \thanks{guendel@bgumail.bgu.ac.il} and
A. B. Kaganovich \thanks{alexk@bgumail.bgu.ac.il}}
\address{Physics Department, Ben Gurion University of the Negev, Beer
Sheva
84105, Israel}
\maketitle
\begin{abstract}

A field theory is proposed where the regular fermionic matter and the dark
fermionic matter are different states of the same "primordial" fermion fields.
In regime of the fermion densities typical for normal particle physics, the
primordial fermions split into three families identified with regular fermions.
When fermion energy density becomes comparable with dark energy density, the theory
allows new type of states. The possibility of such Cosmo-Low Energy Physics
(CLEP) states is demonstrated by means of solutions of the field theory
equations describing FRW universe filled by homogeneous scalar field and
uniformly distributed nonrelativistic neutrinos. Neutrinos in CLEP state
are drawn into
cosmological expansion by means of dynamically changing their own parameters.
One of the features of the fermions in CLEP state is that in the late time
universe their masses
increase as $a^{3/2}$ ($a=a(t)$ is the scale factor). The energy
density of the cold dark matter consisting of neutrinos in CLEP state scales as
a sort of dark energy; this cold dark matter possesses negative pressure and for the
late time universe its equation of state approaches that of the cosmological
constant. The total energy density of such universe is less than it would be in
the universe free of fermionic matter at all. The (quintessence) scalar field
is coupled to dark matter but its coupling to regular fermionic matter
appears to be
extremely strongly suppressed. 

\end{abstract}
\bigskip

  The wish  to combine a possibility to describe an accelerating expansion
for the present day universe with a solution of the cosmic
coincidence problem in the framework of a single consistent model,
was a main motivation of a number of recent attempts (see for
example\cite{VAMP}) to modify the
 particle physics models underlying the quintessence scenarios.
The most
significant   achievement of these models consists in the possibility to
provide both the accelerating expansion and a  resolution of
the coincidence problem without fine tuning.
In spite of this they still have
fundamental problem.
 Although there are  some  justifications for choices of
certain types of
dark matter-dark energy coupling  in the Lagrangian,
there is a necessity to
assume
the absence or extremely strong suppression of the barion matter-dark energy 
coupling. Actually this problem was known from the
very beginning  in the
quintessence models since generically there are no reasons for the absence of  a
direct coupling of the quintessence scalar field $\phi$ to the barion matter.
Such coupling  would be the origin of a long range scalar force because
of the very small mass of the quintessence field  $\phi$.
This ''fifth-force" problem
might be solved if there would be a shift symmetry $\phi\rightarrow\phi +const$
of the action\cite{Carroll}.
 However the quintessence potential itself does not
possess this symmetry. The situation with the ''fifth-force" problem becomes
still more critical in the discussed above models since one should explain now
why the direct quintessence-dark matter coupling is permissible in the Lagrangian
while the same
is forbidden for the barion matter.

Modifications of the particle physics models in
Refs.\cite{VAMP} 
are based on   
 the assumption
that   all the fields of  the fundamental particle theory should be divided into two
large groups:  one describing  detectable particles (ordinary matter) and the other
including dark matter particles.
The main purpose of the presented talk is to demonstrate that there is a field theory which
is able to propose a resolution for the above problems by an absolutely new way:
{\em the dark matter is not introduced as a special type of matter but rather it
appears as the solution of equations of motion describing a new type of states}
 of the (primordial) neutrino field; in other
words, {\em  the dark neutrino matter and the regular neutrino generations
 (electron, muon
and $\tau$ neutrinos) are different states of the same primordial fermion
field}.

This field theory is the Two Measures Theory (TMT) originally built
with 
the aim to solve
the "old" cosmological constant problem\cite{GK1} 
(see also
our other contribution to this section). The
TMT model 
we continue to study in Ref.\cite{nnu} possesses spontaneously broken
global 
scale symmetry\cite{G1}
which includes the shift symmetry $\phi\rightarrow\phi +const$ and it
allows to suggest\cite{GK4,GK5} a simultaneous resolution both of  the
fermion families problem and of the ''fifth-force" problem. {\em The theory
starts from one primordial fermion field for each type of leptons and
quarks}, e.g., in $SU(2)\times U(1)$ gauge theory the fermion content is the
 primordial neutrino and electron fields and primordial $u$ and $d$ quark
fields. It turns out that masses and interactions of fermions
as well as the structure of their contributions to the energy-momentum
tensor depend on the fermion densities.

At a  fermion energy density corresponding
to  {\it normal laboratory particle physics} conditions 
("high fermion density"), the fermion energy-momentum tensor is canonical
and each of the primordial fermions splits into three different states
with different masses (one of these states should be realized via fermion
condensate). These states were identified in  Refs.\cite{GK4,GK5} with   
the mass eigenstates of the fermion
generations, and this effect is treated as the families birth effect.
 The effective  interaction of the dilaton
with the regular fermions of the first two generations appears
to be extremely suppressed.
 In other words, the interaction of the dilaton with matter observable
 in  gravitational experiments is
 practically switched off, and that solves the
''fifth-force" problem.
 
In TMT, physics of fermions at very low densities turns out to be very different
from what we know in normal particle physics. The term {\it very low fermion
density} means here that the fermion
energy density is comparable with the dark energy density. In this
case, in addition to the canonical contribution to the  energy-momentum
tensor, each of the
primordial fermion fields has a noncanonical contribution in the form of a
{\em dynamical fermionic $\Lambda$   term}.

If the fermion
energy density is comparable with the dark energy density then as we
show in Ref.\cite{nnu}, the theory predicts that the
primordial fermion
may not split into  generations and in the FRW universe it can   
 participate in the expansion of the
universe by means of changing its own parameters. We call this effect   
"Cosmo-Particle Phenomenon"  and refer to such states
as  Cosmo-Low Energy Physics (CLEP) states.

As the first step in studying Cosmo-Particle Phenomena, here
we restrict ourselves to the consideration of a simplified cosmological
model where universe  is filled by
a homogeneous scalar field $\phi$ and uniformly distributed
non-relativistic (primordial) neutrinos and antineutrinos in CLEP states.
Such CLEP-neutrino matter is  detectable practically only
through gravitational interaction and this is why it can be regarded
as a model of dark matter. The mass of CLEP-neutrino increases as $a^{3/2}$
where $a=a(t)$ is the scale factor.  This dark matter is also cold one in
the sense that kinetic energy of neutrinos is negligible as compared to their
mass. However due to the dynamical fermionic $\Lambda$   term
generated by neutrinos in CLEP state, this cold dark matter has negative
 pressure and its equation
of state approaches $p_{d.m.}=-\rho_{d.m.}$ as  $a(t)\rightarrow\infty$.
Besides, the energy density of this dark matter scales in a way very
similar to the dark energy which
includes both a cosmological constant and an exponential potential.
So, due to the Cosmo-Particle Phenomena, TMT allows the
universe to achieve both the
accelerated expansion and cosmic coincidence without the need to
postulate the existence of a special sort of matter called "dark matter".
The remarkable feature of such
a Cosmo-Particle solution is that {\em  the total energy density of the
universe
in this case is less than it would be in the universe free
of fermionic matter at all}.
This means that there are two different
vacua: one is the usual vacuum free of the particles, which is
actually a false vacuum, and another one, a true vacuum,
 which could be called "Cosmo-Particle Vacuum" since it is a state
containing neutrinos in CLEP state. Therefore one should expect the possibility
of soft domain walls connecting these two vacua, that may be similar to
soft domain walls studied in Ref.\cite{Schramm}.

In general, the constraint is a seventh order algebraic
equation for $\zeta$ and we should expect that under intermediate
fermion densities conditions, more complex states can exist. One may speculate
that these states (including also some states close to the CLEP states)
may play an important role in the resolution of the halos dark matter
puzzle.

\end{document}